\documentclass{svjour3}
\smartqed  
\usepackage{geometry}
\usepackage{adjustbox}
\usepackage{graphicx}
\usepackage{subfig}
\usepackage{amssymb}
\usepackage{amsmath, amssymb}
\usepackage{booktabs}
\usepackage{tikz}

\usetikzlibrary{arrows,automata,positioning}
\definecolor{mycolor}{rgb}{1,0.647,0}
\tikzset{Node/.style={shape      = circle,
                      scale       = 0.6,
                      fill       = mycolor,
                      text       = black}}
\tikzset{Edge/.style={semithick}}
\tikzset{EdgeBL/.style={bend left,
                       semithick}}
\tikzset{EdgeBR/.style={bend right,
                       semithick}}
\tikzset{Label/.style={text = black}}

\begin{document}

\title{Investigating the Chinese Postman Problem on a Quantum Annealer}

\author{Ilaria Siloi \and Virginia Carnevali \and Bibek Pokharel \and Marco Fornari \and Rosa Di Felice
}
\institute{Ilaria Siloi$^\dagger$  \at
           Department of Physics and Astronomy, University of Southern California, Los Angeles, CA 90089, United States
           \and
           Virginia Carnevali$^\dagger$   \at
           Department of Physics, Central Michigan University, Mt. Pleasant, MI 48859, United States
           \and
           Bibek Pokharel \at
          Department of Physics and Astronomy, University of Southern California, Los Angeles, CA 90089, United States\\
          Center for Quantum Information Science \& Technology, University of Southern California, Los Angeles, California 90089, United States
           \and
          \and
          Marco Fornari$^*$  \at
           Department of Physics and Science of Advanced Materials Program, Central Michigan University, Mt. Pleasant, MI 48859, United States
           \and
          Rosa Di Felice \at
          Department of Physics and Astronomy, University of Southern California, Los Angeles, CA 90089, United States \\
          Center for Quantum Information Science \& Technology, University of Southern California, Los Angeles, California 90089, United States
\and
          $^\dagger$ These Authors contributed equally to this work.\\
          $^*$ Corresponding Author \email{forna1m@cmich.edu}
}

\date{Received: date / Accepted: date}

\maketitle

    \begin{abstract}
The recent availability of quantum annealers has fueled a new area of information technology where such devices are applied to address practically motivated and computationally difficult problems with hardware that exploits quantum mechanical phenomena. D-Wave annealers are promising platforms to solve these problems in the form of quadratic unconstrained binary optimization. Here we provide a formulation of the Chinese postman problem that can be used as a tool for probing the local connectivity of graphs and networks. We treat the problem classically with a tabu algorithm \textcolor{black}{and simulated annealing}, and using a D-Wave device. \textcolor{black}{The efficiency of quantum annealing with respect to the simulated annealing has been demonstrated using the optimal time to solution metric.} We systematically analyze computational parameters associated with the specific hardware. Our results clarify how the interplay between the embedding due to limited connectivity of the Chimera graph, the definition of logical qubits, and the role of spin-reversal controls the probability of reaching the expected solution.
\end{abstract}
   \keywords{D-Wave, Quantum annealing, QUBO, routing problems}

\section{Introduction}
Since their proposal by Kadowaki and Nishimori \cite{Kadowaki_PhysRevE1998}, quantum annealers (such as D-Wave machines) have advanced to the point that there is now a community of users whose goal is mainly to apply adiabatic quantum optimization (AQO) to a diverse set of computational problems in fields ranging from {materials \cite{venturelli2015quantum} and biological properties \cite{DiFelice,perdomo2012finding} to machine learning \cite{li2019unconventional,neven2008image}, fault detection \cite{faultdetection_ICT2016} and optimization \cite{Neukart,Bian_2019,Stollenwerk_IEEE2020,SHOWALTER2014395,8656874,Gomes2016OptimizationOR}}.
Adiabatic quantum computation has been extensively reviewed \cite{albash2018adiabatic,Das_RevModPhys2008,mcgeoch_book2014,aharonov2008adiabatic}, as well as hardware/software aspects of D-Wave quantum annealers \cite{Dwave_Nature2011,Raymond_ICT2016,HardwareDWave_IEEE2014,boixo2014evidence,albash2015reexamining}.
However, it remains important to expand the library of applications of quantum annealing for several reasons. First, these problems are stepping stones on the way to solving practical problems that may be beyond the reach of classical computation. Further, by comparing these algorithms with their classical counterparts one can probe the computational reach of quantum devices. Lastly, they enable both the providers and the users to identify optimal modes of operation and necessary improvements for the currently available machines.

AQO proceeds from an initial Hamiltonian $H_0$ to a final Hamiltonian $H_1$ whose ground state encodes the solution of the computational problem under consideration \cite{Santoro_Science2002}. The evolution is controlled by $t \in [0,T]$ (possibly with $T\rightarrow \infty$) through two monotonic functions $A(t)$ and $B(t)$ such as $A(T)=0$ and $B(0)=0$: $$H(t)=A(t)H_0+B(t)H_1.$$ Starting from the ground state of $H_0$, the adiabatic theorem guarantees that the quantum state will remain in the ground state of $H(t)$, under Schr\"{o}dinger evolution, provided that the Hamiltonian is varied slowly enough \cite{Kadowaki_PhysRevE1998}.
In D-Wave systems, the initial configuration is given by the ground state of $H_0=-\sum_{i}h_i \sigma^x_i$ ($\sigma^\alpha_i$ are the Pauli operators for spin $i$ in the direction $\alpha$) with all the spins aligned along the $x$-direction. During the adiabatic process those spins start to interact {\it \`a la} Ising, $$H_1=\sum_{ik}J_{ik}\sigma^z_i\sigma^z_k+\sum_{i}h_i\sigma^z_i,$$ according to the specific choice of the parameters $J_{ik}$ and $h_i$.
Since $[H_0,H_1] \ne 0$,  the quantum evolution is not trivial.
At the end of each adiabatic cycle, a reading of the spin configuration in the $z$-basis provides a classical sample with a specific energy ($[\sigma^z_i,H_1]=0$).
Repeated measurements allow to extract the probability distribution for the solution.
AQO mimics classical simulated thermal annealing but uses  quantum superposition and tunneling instead of thermal fluctuations in order to reach a global minimum \cite{Denchev_PhysRevX2016}. While D-Wave has been successfully applied to solve many difficult problems, the advantages of D-Wave system to analyze optimization problems over classical algorithms are not clear; noise and decoherence play an important role and performance depends on parameters that are not easily controllable \cite{mishra2018,Tameem_PhysRevX2018,parekh2016benchmarking,karimi2012investigating}.

In this paper, we use D-Wave to analyze the undirected Chinese postman problem (CPP) \cite{Garey}
originally formulated in 1962 by the mathematician Kwan Mei-ko \cite{CPP}. The CPP
involves finding the ``length'' of the shortest closed path traveling across all edges of the network at least once. From a practical point of view the CPP problem is of interest in many situations where something or someone has to periodically traverse or inspect every link in a network, e.g. parallel programming, security patrolling, school bus route, etc. Modifications of routing problems may be used to model defects and transport in solids, which motivates us to choose CPP problem for the exploration of quantum computation in materials science. \textcolor{black}{Besides all the possible applications, it is worthy to underline that none of the problems in the CPP class has been yet solved on a quantum annealer. The idea of this work is to implement the easiest CPP problem as a starting point towards developing generalizations to more complicated situations.}
There are several algorithms that solve the CPP \cite{Pearn,Ahr,Zhang,Eiselt}. This paper introduces a quadratic unconstrained binary optimization (QUBO) formulation of the CPP, which can be programmed into a quantum annealer. In addition, we run our algorithm on D-Wave 2X, detail the parameters that control the quality of the results, and exploit the solutions to probe features of the network topology.

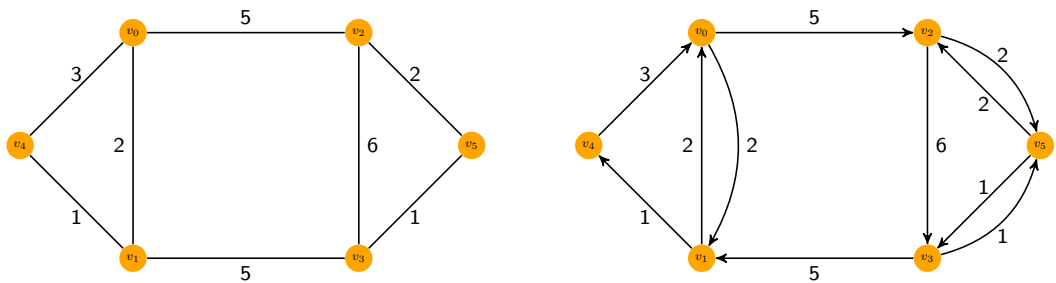
\begin{figure*}[hptb]
\centering
\begin{minipage}{.5\textwidth}
\centering
\begin{tikzpicture}
  \node[Node] (n6) at (3,0) {$v_5$};
  \node[Node] (n4) at (1.5,1.5)  {$v_2$};
  \node[Node] (n5) at (1.5,-1.5)  {$v_3$};
  \node[Node] (n1) at (-3,0) {$v_4$};
  \node[Node] (n2) at (-1.5,1.5)  {$v_0$};
  \node[Node] (n3) at (-1.5,-1.5)  {$v_1$};

  \path[every node/.style={font=\sffamily\small}]
  (n1) edge[Edge] node[above]{3} (n2)
  (n1) edge[Edge] node[below]{1} (n3)
  (n2) edge[Edge] node[above]{5} (n4)
  (n3) edge[Edge] node[below]{5} (n5)
  (n2) edge[Edge] node[left]{2} (n3)
  (n4) edge[Edge] node[right]{6} (n5)
  (n4) edge[Edge] node[above]{2} (n6)
  (n5) edge[Edge] node[below]{1} (n6);

\end{tikzpicture}
\end{minipage}%
\begin{minipage}{.5\textwidth}
\centering
\begin{tikzpicture}[->,>=stealth']
  \node[Node] (n6) at (3,0) {$v_5$};
  \node[Node] (n4) at (1.5,1.5) {$v_2$};
  \node[Node] (n5) at (1.5,-1.5) {$v_3$};
  \node[Node] (n1) at (-3,0) {$v_4$};
  \node[Node] (n2) at (-1.5,1.5)  {$v_0$};
  \node[Node] (n3) at (-1.5,-1.5)  {$v_1$};

  \path[every node/.style={font=\sffamily\small}]
  (n1) edge[Edge] node[above]{3} (n2)
  (n3) edge[Edge] node[below]{1} (n1)
  (n2) edge[Edge] node[above]{5} (n4)
  (n5) edge[Edge] node[below]{5} (n3)
  (n2) edge[EdgeBL] node[right]{2} (n3)
  (n4) edge[Edge] node[right]{6} (n5)
  (n4) edge[EdgeBL] node[above]{2} (n6)
  (n6) edge[Edge] node[above]{1} (n5)
  (n5) edge[EdgeBR] node[below]{1} (n6)
  (n3) edge[Edge] node[left]{2} (n2)
  (n6) edge[Edge]node[below]{2} (n4);

\end{tikzpicture}
\end{minipage}%
\caption{{Left panel:} The network $G(E,V,w)$ used to illustrate the solution of the CPP. {Right panel:} One possible solution path of the CPP in this specific network.}
\label{Ex}
\end{figure*}

\section{The Chinese Postman Problem}
\label{CPPSec}
The CPP is modeled using a network
$G(V,E,w)$ (with $V=\{v_1, v_2, \dots\}$ being the set of all the nodes, $E=\{(v_1,v_2), \dots, (v_i,v_j),\dots\} \subset V \times V$ the set of all the edges, and $w: E \rightarrow \mathbb{R}$ a mapping assigning a ``length'' $w_{ij}$ to each edge). The goal is to find the shortest closed  path length that crosses all the elements of $E$ at least once.
While the undirected CPP and the directed CPP can be solved in polynomial time, common generalizations (e.g. the mixed CPP and the rural CPP) are NP-hard \cite{CPPApplications_2005}.
The CPP admits a solution {if and only if} there exists at least one Eulerian cycle, i.e. a cycle that crosses each edge exactly once \cite{Euler,Hedetniemi}. A finite graph contains zero odd degree nodes (Eulerian graph) or an even number (non-Eulerian graph).
In case of non-Eulerian network topologies, the algorithm dictates that extra paths linking odd degree nodes must be added to guarantee the existence of an Eulerian cycle; in other words the algorithm allows for the edges to be crossed more than once. Then, the shortest extra path has to be chosen. \textcolor{black}{The two cases above are solved exactly}.\\
In the network of Fig.\,\ref{Ex}a, for instance, there are four nodes of odd degree, $V_O=\{v_0,v_1,v_2,v_3\} \subset V$ ($|V_O| = d = 4$) and three possibilities for the extra paths
\begin{equation}
  \begin{split}
  m(\pi_1)=\quad &W(v_0,v_1)+W(v_2,v_3)=2+3=5 \\
  m(\pi_2)=\quad  &W(v_0,v_2)+W(v_1,v_3)=5+5=10 \\
  m(\pi_3)=\quad &W(v_0,v_3)+W(v_1,v_2)=7+7=14.
  \label{pi}
\end{split}
\end{equation}
The ``length'' of each extra path between pairs of odd degree nodes is determined by computing the minimum across all the possible paths in $G(V,E,w)$ , e.g. $W(v_0,v_3$) is the minimum between $w(v_0,v_1)+w(v_1,v_3)=7$, $w(v_0,v_2)+w(v_2,v_3)=11$, $w(v_0,v_5)+w(v_5,v_1)+(v_1,v_3)=9$, etc. The shortest ``length'', which is the solution of the CPP problem, is given by the sum of all the arcs' lengths plus the shortest extra path, $$l_T(G) = \sum_{e \in E} w(e) + W(v_0,v_1)+W(v_2,v_3)= 25+5.$$ The path corresponding to the solution of the CPP is shown  Fig. \ref{Ex}b. Note that the path is not unique; at least a second path exists by the inversion of all the edge directions (since we are dealing with an undirected CPP).

Goodman and Hedetniem \cite{Goodman} demonstrated that given a finite undirected network $G(V, E,w)$ with $d$  odd degree nodes, it is always possible (1) to find $(d-1)!!$ perfect matchings $\pi_\alpha$
of pairs of odd degree nodes (as in Eq. \ref{pi}), (2) to assign to each $\pi_\alpha$ a minimal path ``length" $m(\pi_\alpha)$, and (3) to determine the solution of the CPP by using
\begin{equation}
  l_T(G) = \sum_{e \in E} w(e) + \min_{\alpha}m(\pi_{\alpha}) = \sum_{e \in E} w(e) + M_{min}
  \label{mg}
\end{equation}

\section{QUBO for the CPP}
In quantum annealing, optimization problems are encoded as Ising Hamiltonians whose ground state configuration is a binary string solution to the corresponding problem. Ising Hamiltonian can be mapped to quadratic unconstrained binary optimization (QUBO). Finding the ground state of the Ising then is equivalent to the minimization of a quadratic form in $\mathbb{Z}_2^{2N}$ \cite{Laughhunn}.

\textcolor{black}{Here, we are going to derive for the first time the QUBO formulation for the CPP problem.} In order to do that, the objective function $m(\pi_\alpha)$ has to be expressed as binary integer problem (BIP, for a binary variable, $x=x^2$). Given an undirected network $G(V,E,w)$ and $V_O \subset V$ with $d$ nodes of odd degree, the possible paths can be represented by the variable $x_{ij}$ where $i,j \in \{0,..,d-1\}$ and  $i\neq j$ such as:
\begin{equation}
x_{ij}=
 \begin{cases}
   1 \quad \text{if node}\, i \, \text{is paired with node}\, j\\
   0 \quad \text{otherwise.}
 \end{cases}
\end{equation}
Thus, a binary vector $\mathbf{x}\in \mathbb{Z}_{2}^{d(d-1)}$ of decision variables for the QUBO can be used to search for the global shortest ``length" by minimizing the sum of all the shortest distances of all the permutations of nodes of odd degree.
\begin{equation}
  F(\mathbf{x})=\sum_{i\ne j}^{d-1}W_{ij}x_{ij},
  \label{F}
\end{equation}
where $W_{ij} = W(v_i,v_j)$ is the ``length" of the minimum path between the odd degree nodes $i$ and $j$ by taking into account all the network features as discussed in Sec. \ref{CPPSec}. In the case of the CPP, $F(\mathbf{x})$ is subject to two constrains which guarantee that the ordered pairs of nodes are unique and that each node is counted only once in each subset $\pi_\alpha$. $P_1(\mathbf{x})$
penalizes double counting of pairs of nodes in $V_O$ and  $P_2(\mathbf{x})$ penalizes whenever the combination of the pairs is not legal (see the Supplementary Material for details).
Overall, the quadratic function to be optimized is
\begin{equation}
  Q(\mathbf{x})=\sum_{i\ne j}^{d-1}W_{ij}x_{ij}^2+p P_1(\mathbf{x})+p P_2(\mathbf{x})
  \label{Qubo}
\end{equation}
with
\begin{equation}
  P_1(\mathbf{x})=\sum_{i\ne j}^{d-1}\left(1-\sum_{j=0}^{d-1}(x_{ij}+x_{ji})\right)^2
  \label{P1}
\end{equation}
and
\begin{equation}
  P_2(\mathbf{x})=\sum_{i\ne k; j \ne k}^{d-1}(x_{ik}x_{jk}+x_{ki}x_{kj})
  \label{P2}
\end{equation}
for some constant $p \geqslant d$.

For the network shown in Fig. \ref{Ex} all the ordered pair of vertices of odd degree can be easily listed:
\begin{equation*}
  \begin{split}
   &\{(v_0,v_1),(v_0,v_2),(v_0,v_3),(v_1,v_0),(v_1,v_2),(v_1,v_3),\\
   &(v_2,v_0),(v_2,v_1),(v_2,v_3),(v_3,v_0),(v_3,v_1),(v_3,v_2)\}
 \end{split}
\end{equation*}
and the representing vector is $\mathbf{x}=(x_{01},x_{02},x_{03},x_{10},x_{12},x_{13},x_{20},x_{21},x_{23},x_{30},x_{31},x_{32})\in \mathbb{Z}_{2}^{12}$. The legal combinations of pairs are represented by $\pi_{\alpha}$ with $\alpha = 1,2,3$.
The coefficients of the matrix representing the quadratic form are determined using $p=8$ and are shown in Table \ref{QuboEx}. The minimum of $Q(\mathbf{x})$ is given by
\begin{equation*}
  \mathbf{x_*}=(1,0,0,0,0,0,0,0,1,0,0,0),
  \label{ExSol}
\end{equation*}
meaning that the legal partition of pairs formed with odd degree nodes is $\{(v_0,v_1),(v_2,v_3)\}$. The shortest paths are shown in Fig. \ref{Ex}b. The minimum sum of pairs of nodes with odd degree  is given by $Q(\mathbf{x_*})$.

\begin{table*}[hptb]
 \centering
  \begin{tabular}{c|ccc|ccc|ccc|ccc} \toprule
    {Pairs} & {$x_{01}$} & {$x_{02}$} & {$x_{03}$} & {$x_{10}$} & {$x_{12}$} & {$x_{13}$} & {$x_{20}$} & {$x_{21}$} & {$x_{23}$} & {$x_{30}$} & {$x_{31}$} & {$x_{32}$} \\ \midrule
    $x_{01}$  & -12 & 16 & 16 & 48  & 16 & 16 & 16 & 16 & 0   & 16 & 16 & 0  \\
    $x_{02}$  & 16  & -6 & 16 & 16  & 16 & 0  & 48 & 16 & 16  & 16 & 0  & 16 \\
    $x_{03}$  & 16  & 16 & -2 & 16  & 0  & 16 & 16 & 0  & 16  & 48 & 16 & 16 \\ \midrule
    $x_{10}$  & 48  & 16 & 16 & -12 & 16 & 16 & 16 & 16 & 0   & 16 & 16 & 0  \\
    $x_{12}$  & 16  & 16 & 0  & 16  & -2 & 16 & 16 & 48 & 16  & 0  & 16 & 16 \\
    $x_{13}$  & 16  & 0  & 16 & 16  & 16 & -6 & 0  & 16 & 16  & 16 & 48 & 16 \\ \midrule
    $x_{20}$  & 16  & 48 & 16 & 16  & 16 & 0  & -6 & 16 & 16  & 16 & 0  & 16 \\
    $x_{21}$  & 16  & 16 & 0  & 16  & 48 & 16 & 16 & -2 & 16  & 0  & 16 & 16 \\
    $x_{23}$  & 0   & 16 & 16 & 0   & 16 & 16 & 16 & 16 & -10 & 16 & 16 & 48 \\ \midrule
    $x_{30}$  & 16  & 16 & 48 & 16  & 0  & 16 & 16 & 0  & 16  & -2 & 16 & 16  \\
    $x_{31}$  & 16  & 0  & 16 & 16  & 16 & 0  & 0  & 16 & 16  & 16 & -6 & 16  \\
    $x_{32}$  & 0   & 16 & 16 & 0   & 16 & 16 & 16 & 16 & 48  & 16 & 16 & -10 \\ \bottomrule
  \end{tabular}
  \caption{QUBO matrix for CPP on the network in Fig.\,\ref{Ex}. The penalty constant is chosen to be equal to the number of nodes\textcolor{black}{+2} ($p=8$). {Details on the derivation of the matrix elements $Q_{ij}$ can be found in the SM, Sec.\ 1.}}
  \label{QuboEx}
\end{table*}

\section{Computational Methods}\label{comp}

We first solve several cases of the CPP classically using the software {\tt qbsolv} \cite{Qbsolv_new}, a solver that finds the minimum value for QUBO problem using a metaheuristic tabu search algorithm \cite{Gendreau2003}.
We then implement our formulation on D-Wave 2X following a protocol that facilitates comparisons with the classical solutions and provide insight on the parameters of the calculations. \textcolor{black}{In order to estimate the efficiency of the quantum annealing we also compared with classical simulated annealing (SA) following the work of Albash and Lidar \cite{Tameem_PhysRevX2018}}.

In D-Wave systems, physical qubits are arranged in a Chimera graph topology; in order to be able to represent   the QUBO, a set of linked (logical) qubits must be defined \cite{S_Humble_2014}. Physical qubits are organized in chains to ``simulate" the logical qubits;
this is known as a minor embedding \cite{min_emb}. D-Wave's API provides a function called {\tt minorminer}  which heuristically searches for the optimal (minor)embedding \cite{Cai_2014,Choi2011}. {\tt minorminer} minimizes the length of each embedding chain, as shorter chains are less prone to breaking into domains where physical qubits have opposite spin orientations.
The choice of intra-chain coupling ($J_F$) is important as it affects the time-dependent energy spectrum in the adiabatic evolution and determines the ability of the chain to act as a single variable. $J_F$ couplings should be strong enough to avoid chain-breaking without dominating the dynamics. Tight bounds on these conditions are derived in Ref. \cite{Choi2011}.
In addition, due to the dynamic range and the precision of the hardware control \cite{Barahona}, the representation of the QUBO in terms of the Ising Hamiltonian parameters depends on the relative scale of $J_{ij}$ and $h_i$.

Our protocol on D-Wave involves: (1) selecting a network to study the CPP, (2) scaling appropriately the entries of the QUBO matrix, (3) embedding the specific QUBO topology on the Chimera graph using {\tt minorminer}, (4) assessing the quality of different embeddings, (5) optimizing intra-chain coupling ($J_F$), (6) comparing the results with {\tt qbsolv} when possible.

\begin{figure*}[hptb]
  \centering
\includegraphics[scale=0.22]{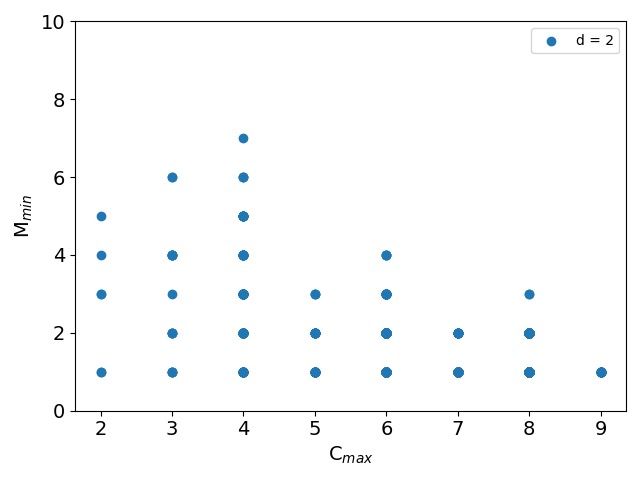}
\includegraphics[scale=0.22]{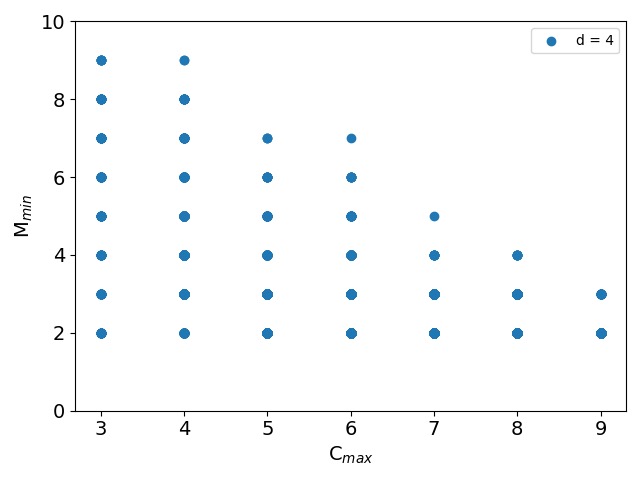}
\includegraphics[scale=0.22]{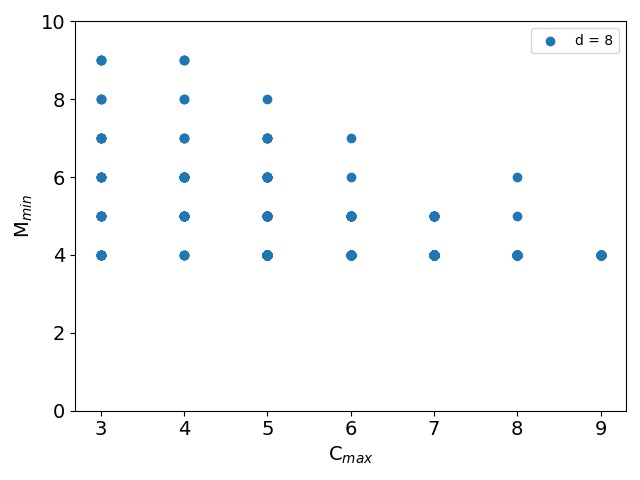}
  \caption{$M_{min}$ as a function of the largest degree of nodes in the graph (we used $w_{ij} = 1\ \forall\  i,j$ to simplify the analysis).
  The results were obtained with a sample of 10,000  randomly generated graphs with $n=10$. Results are divided according to the number of nodes with odd degree ($d$) in the graph. We report only three representative subsets in this figure. {Left panel:} 875 graphs with $d=2$. {Middle panel:} 4,025 graphs with $d=4$. {Right panel:}  809 graphs with $d=8$.}
  \label{max_all}
\end{figure*}

\section{Classical Results by {\tt qbsolv}}\label{qbsolv}

In order to test the QUBO formulation and establish references for the D-Wave calculations, we randomly generate 10,000 non-Eulerian networks of order $n$ with variable number of nodes with odd degree $d \geqslant 2$ and variable size (the number of edges specified in $E\subset V\times V$). 

\begin{figure*}[hptb]
  \centering
  \includegraphics[scale=0.45]{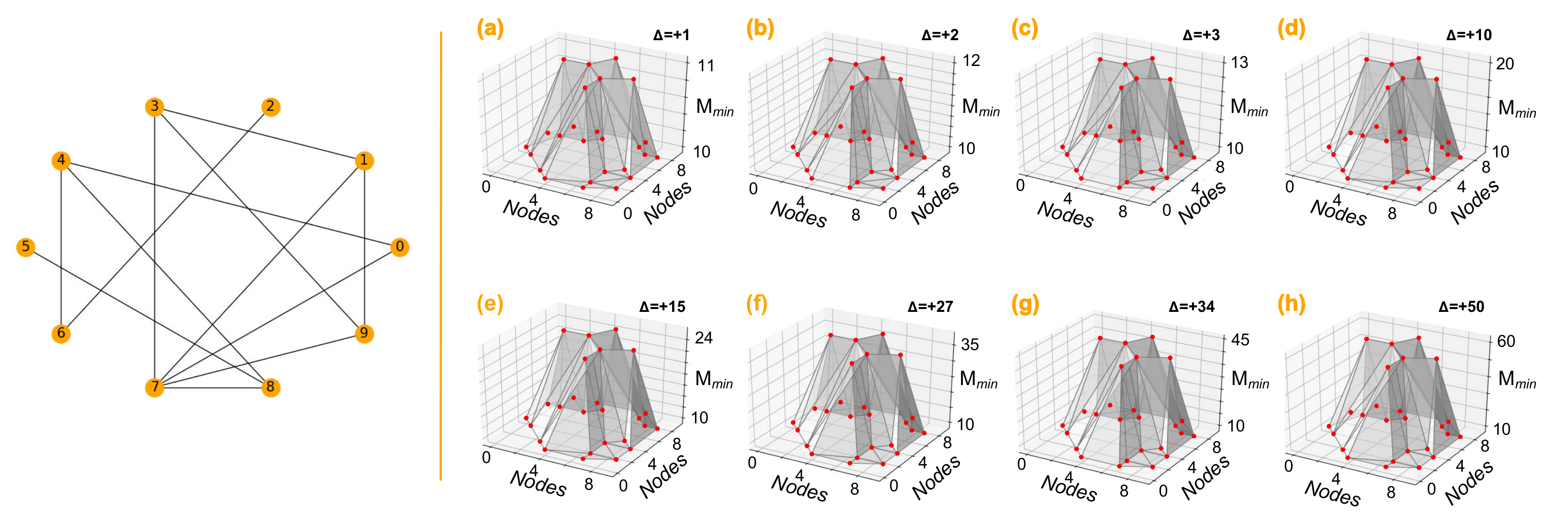}
  \caption{{Left panel:} Visual representation of the graph $G(E,V,w=1)$ with $n$=10 nodes and $d$=8.
   {Right panel:} value of $M_{min}(i,j)$ as function of the $(i,j)$ position of the defect in the adjacency matrix. Each plot corresponds to a different value ($\Delta$) for the defected edge.}
  \label{10N}
\end{figure*}
As expected, longer CPP paths ($M_{min}$, see Eq.\,\ref{mg}) are associated with the presence of nodes with small degree, conversely if there are large degree nodes, the value of $M_{min}$ is reduced to the minimum $M_{min}=d/2$. Notably, the result of the optimization problem depends on the degree of all the nodes in the network.
In Fig.\ref{max_all}, we report $M_{min}$ as function of the maximum degree ($c_{MAX}$) for graphs with $n=10$ where edge weights are all set to unity, $w_{ij} = \epsilon=1\ \forall \ i,j$. The CPP solution is sensible to the degree of the vertices suggesting that the CPP can be used to explore the degree of the nodes in a graph.
 In order to gain additional insight, we introduce multiple random ``defects" in the network by varying one or more $w_{ij}$. In particular,  we analyze $M_{min}$ in the case of $n=10,\,14,\,17$ (see Tab.\ref{TabGraphs}) by setting \textcolor{black}{from one to three}edges to $\epsilon + \Delta$, with $\Delta=1,2,3,10,15,27,34,50$. \textcolor{black}{Given a graph, the CPP has been solved for all the possible arrangements of one, two, and three defected edges in the graph. A total of 1,000 random graphs has been investigated with all the $\Delta$ values. \textcolor{black}{Our intention is to explore ways to use the CPP algorithm to characterize defected networks.}}
 \begin{table}[hptb]
  \centering
   \begin{tabular}{c|c|c|c|c} \toprule
     {$n$} & $d$ & $c_{MAX}$ & $c_{min}$ & $c_1$ \\ \midrule
     10  & 8 & 3 & 1 & 2  \\
     14  & 4  & 5 & 1 & 2  \\
    17  & 4  & 8 & 2 & 0  \\ \bottomrule
   \end{tabular}
   \caption{Main features of the graphs: number of vertices ($n$), number of nodes with odd degree ($d$), max degree ($c_{MAX}$) and min degree ($c_{min}$) among all nodes, number of nodes with degree equal to 1, ($c_1$).}
   \label{TabGraphs}
 \end{table}

For $n=10$, a reference graph with 8 nodes with odd degree that gives $M_{min}=10$ was chosen to illustrate the effect (see Fig.\,\ref{10N} left panel).
The right panel of Fig.\,\ref{10N} shows the value of $M_{min}$ as function of the position of the defect in the $(i,j)$ element of the the adjacency matrix of the graph ($M_{min}(i,j)$). Due to the introduction of a single defect, 6 distinct peaks arise regardless the  $\Delta$ value. Despite the difference in the absolute values of the maximum $M_{min}$ with  $\Delta$, features and shapes are preserved. \textcolor{black}{Since the adjacency matrix $A$ is, by definition, symmetric,} only 3 distinct edges, $(v_2,v_6),(v_5,v_8),(v_4,v_6)$, are crucial for the CPP. Two different behaviors can be observed: (1) the edges $(v_2,v_6),(v_5,v_8)$ link nodes with degree one ($v_2$ and $v_5$) or (2) $M_{min}$ is maximum for the edge $(v_4,v_6)$ is associated to a node of degree 2 ($v_6$). In general, $M_{min}$ reaches a maximum value when the postman is forced to travel twice of the edge with increased ``length" (defect)  when nodes with small degree are present.
Similar behavior is observed in the case of two defects, while for three defects in a small graph is impossible to extract information regarding the local topology (nodes' degree).

\begin{figure*}[hptb]
   \centering
   \includegraphics[scale=0.45]{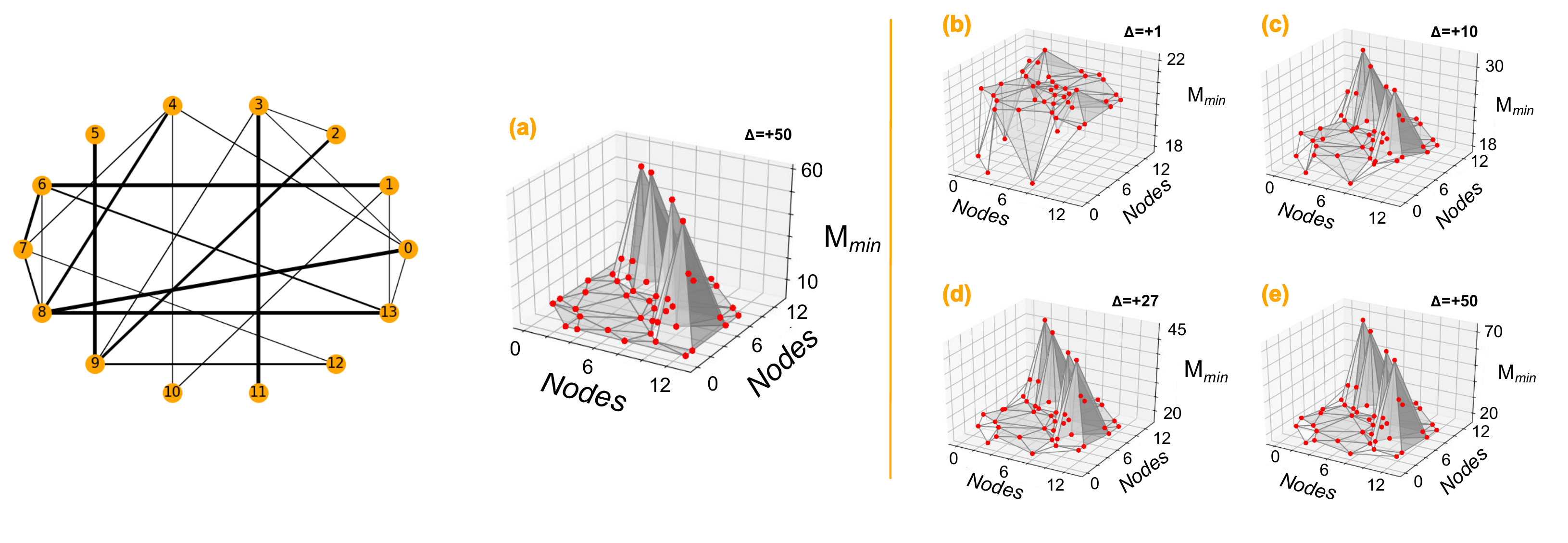}
   \caption{{Left panel:} Visual representation of the network $G(E,V,w)$ with order $n=14$, four odd degree nodes ($d=4$ $\{v_1,v_5,v_8,v_{11}\}$: the weights $w_{ij} \in \{1,\dots,5\}$ are shown with line thickness. {(a)} $M_{min}(i,j)$ is represented as in Fig.\ \ref{10N} for $\Delta=0$. {Right panel:} Plots illustrate the effect of introducing defects of magnitude $\Delta$ on the edge $(v_i,v_j)\in E$.}
   \label{14N_net}
 \end{figure*}

  Finally, we analyze the case of networks with $n$=10, 14, 17 with random $w_{ij}$. One specific case of a network with $n=14$, $d=4$, and $c_1=2$ ($c_1$ is the number of vertices with degree equal to one) is shown in Fig.\ \ref{14N_net}. In case of a large defect ($\Delta \gg w_{ij}\ \forall\ i,j$) the plots in Fig.\ \ref{14N_net} resemble the ones in Fig.\ \ref{10N}.
 The edge weights take values in the range $w\in\{1,\dots,5\}$. For $\Delta \geqslant 27$ the $M_{min}$ representation (we used the same format as in Fig.\ \ref{10N}) is equivalent to the one of the graph (Fig.\ref{14N_net}a). This is reasonable, if the defected edge is linked to one of the two nodes with degree one, it has to be included in the $M_{min}$ computation. Overall, for arbitrary networks assessing $M_{min}(i,j,\Delta)$ is an effective way to detect nodes of degree one. The appropriate $\Delta$ value required to access the local connectivity depends on $c_{MAX}$, $d$, and the size of the network.

\section{Results by Quantum Annealing (D-Wave)}

\begin{figure*}[hptb]
  \centering
  \includegraphics[scale=0.077]{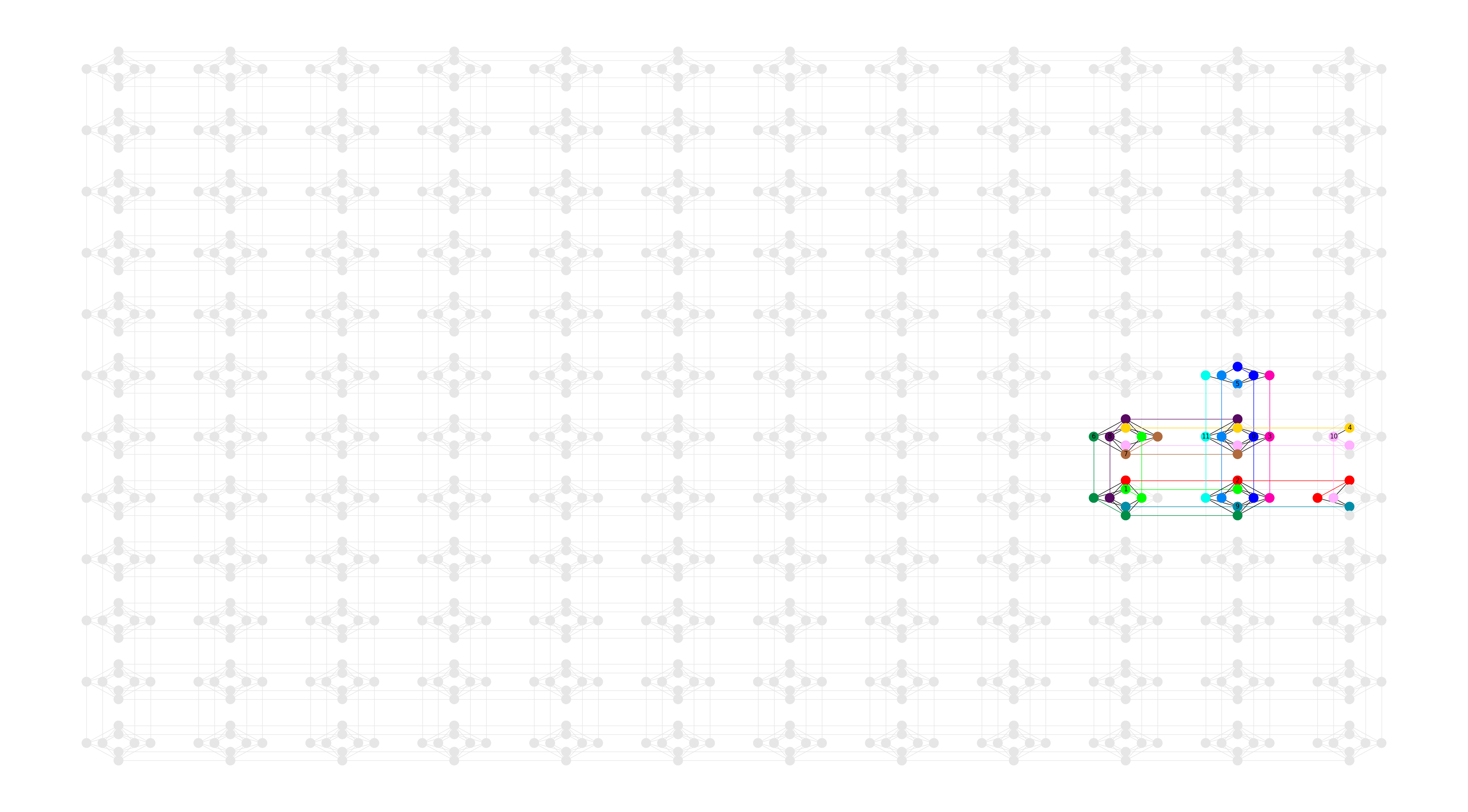}
  \includegraphics[scale=0.077]{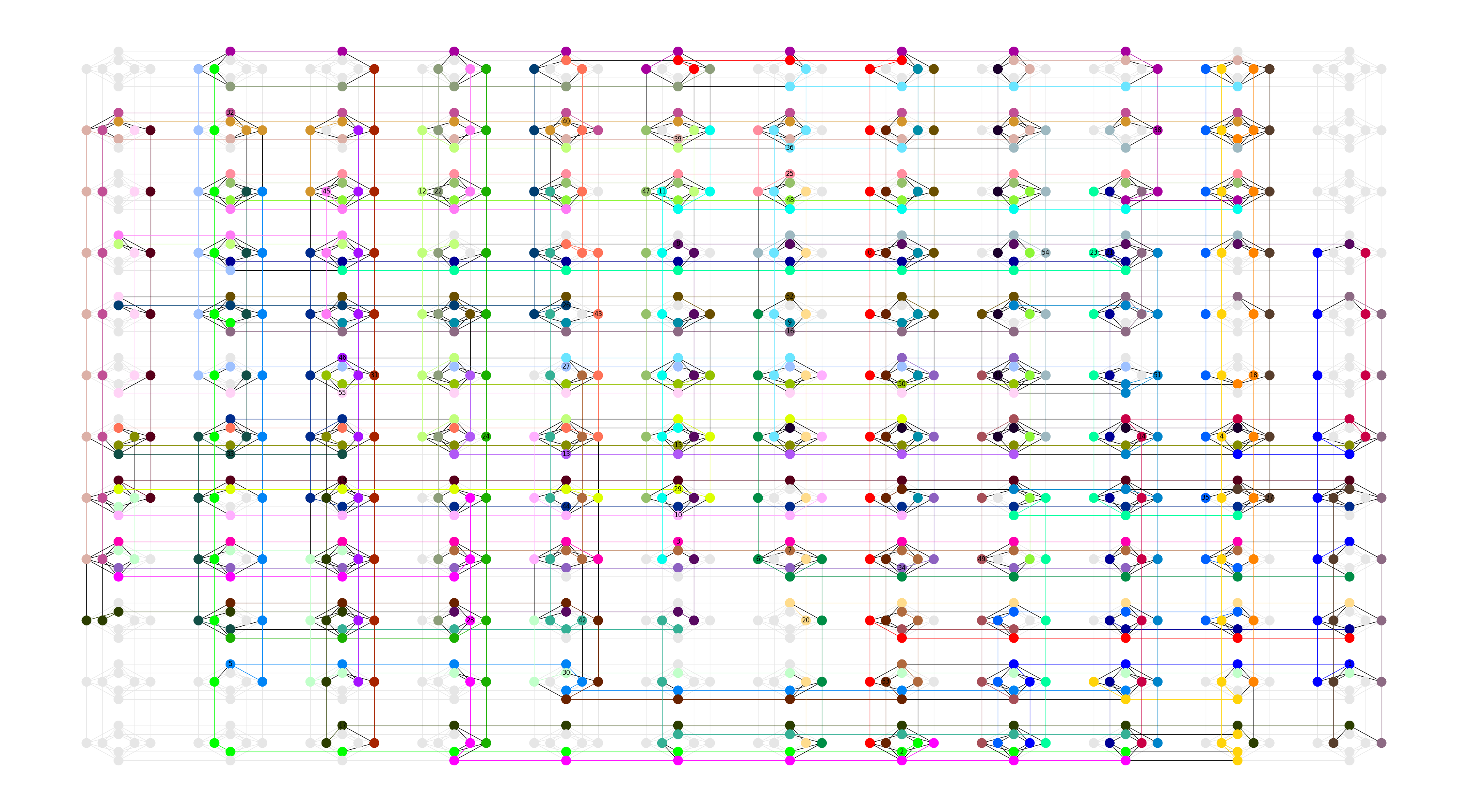}
  \caption{Map of two of the ``optimal" embeddings on the whole Chimera graph for case of a QUBO with $d=4$ (left) and $d=8$ (right). Different colors highlight different spin-chains, namely different logical qubits. The case for $d=8$ has long chains and uses 902 out of 12 $\times$ 12 $\times$ 8 physical qubits in D-Wave 2X. This makes very difficult to control the noise induced by chain breaking.}
  \label{emb}
\end{figure*}

Using D-Wave 2X, we analyze the quality of the CPP solution by changing embedding parameters (see Sec.\ \ref{comp}), the intra-chain coupling ($J_F$), and the scaling of the QUBO to fit the dynamic range of the hardware.
Initially, we select graphs ($w_{ij}=1\ \forall i,j$) with increasing number of odd degree nodes ($d=2,4,6,8$, intuitively pointing to the complexity of the specific CPP), rescale the QUBO matrix elements (using {\tt autoscale}), and embed the CPP problem on $12\times 12\times 8$ physical qubits (minus 54 malfunctioning physical qubits) of the D-Wave Chimera.
The ``quality" of the embeddings is assessed using the probability of reaching the desired ground state (P$_{gs}$) and the performance of DWave 2X is assessed by computing the time to get the solution with probability 99\%:
$$
\mbox{T}_{99}=\dfrac{ln(1-0.99)}{ln(1-\mbox{P}_{gs})} T
$$
where T is the annealing time.
Even after selecting the embedding (see for instance Fig.\ \ref{emb} for the cases $d=4$ and $d=8$), the P$_{gs}$ depends significantly on the choice of the intra-chain coupling $J_F$ \cite{venturelli2015quantum,Stollenwerk_IEEE2020} \textcolor{black}{and the annealing time T. Because the best T is different from one embedding to another, it is not possible to select a T that allows top performance for all the embeddings. According to this, we have evaluated $\mbox{T}_{99}$ using the default T=$20\mu s$ ($\mbox{T}^{A}_{99}$) and the optimal annealing time for the best embedding at a given size $d$ of the problem ($\mbox{T}^{B}_{99}$). The performance of D-Wave 2X has been compared also with simulated annealing as implemented in D-Wave Ocean SDK \cite{Ocean}. Following Albash and Lidar \cite{Tameem_PhysRevX2018}, the time to solution for SA reads as:
$$
\mbox{TTS}=N^2\dfrac{ln(1-0.99)}{ln(1-\mbox{P}_{gs})}\tau_{s}n_{s}
$$
where $N$ is the size of the graph (see Supporting Material for further clarifications), $n_{s}$ is the number of sweeps, and $\tau_{s}=1/f_{SA}$ the time required to perform a single sweep ($f_{SA}$ is the number of spin updates per unit time). Because the SA algorithm performs one spin flip per time step and the clock rate of our cpu is $2.4$ GHz, $f_{SA}=2 \,\text{ns}^{-1}$. The chosen SA annealing schedule $\beta$ is comparable to the one of D-Wave 2X.}

\begin{table*}[hptb]
\centering
\begin{tabular}{c|c|c|c|c|cccc} \toprule
 {$d$} & \shortstack{Logical \\ qubits} &  \shortstack{Number of \\ Q$_{ij}$ terms} & \shortstack{Physical \\ qubits } & \shortstack {P$_{gs}$} (\%) & \shortstack {T$^{A}_{99}$ (s)} & \shortstack {T$^{B}_{99}$ (s)} &  \shortstack{ TTS (s)}\\ \midrule
    2  & 2  &  2  & 2   & 99.99 & 9.7e-6 & 7.1e-6 & 4.0e-4 &\\
   4  & 12 & 54 & 44 & 87.83 & 4.37e-5 & 2.0e-5 & 7.0e-3 &\\
   6  & 30 & 256 & 248 & 51.07 & 1.29e-4 & 2.3e-4 & 1.8e-2 &\\
   8  & 56 & 700  & 864 & 0.21& 4.38e-2 & 1.8e-2 & 7.9e-2 &\\ \bottomrule
\end{tabular}
\caption{Parameters and performance of the ``optimal" embedding for the CPP cases studied with D-Wave 2X: number of nodes with odd degree ($d$), number of logical qubits (equal to the number of variables in the optimization problem), number of terms in the QUBO, number of physical qubits required in the embedding, best probability of finding the correct ground state P$_{gs}$, \textcolor{black}{time to solution with T=20 $\mu s$ (T$^{A}_{99}$), optimal time to solution T (T$^{B}_{99}$), time to solution of the simulated annealing TTS.} The embeddings were generated with {\tt minorminer} and chosen according to the criteria discussed in the text.}
\label{TabDW}
\end{table*}

Tab.\ \ref{TabDW} reports the information for the ``optimal" embeddings (largest P$_{gs}$) for each problem size ($d$) we considered. For $d=2$ physical and logical qubits are the same so the D-Wave results match the classical results consistently (P$_{gs}$=99.97\%, see Tab.\ \ref{TabDW}). However, for $d\geqslant 4$, embeddings become necessary.
When the chain alignment in an embedding breaks, D-Wave automatically performs a majority vote to assign a value to the corresponding logical qubit: we report in the Supplementary Material (SM Tab.\ 1) detailed information on the topology of the embeddings as well as P$_{gs}$ and the best time to solution (T$_{99}$) counting all chains regardless possible loss of intra-chain alignment. Here we only consider the solution without broken chains, and do not perform any post-processing of the D-Wave solutions.

 \begin{figure*}[hptb]
   \centering
   \includegraphics[scale=0.5]{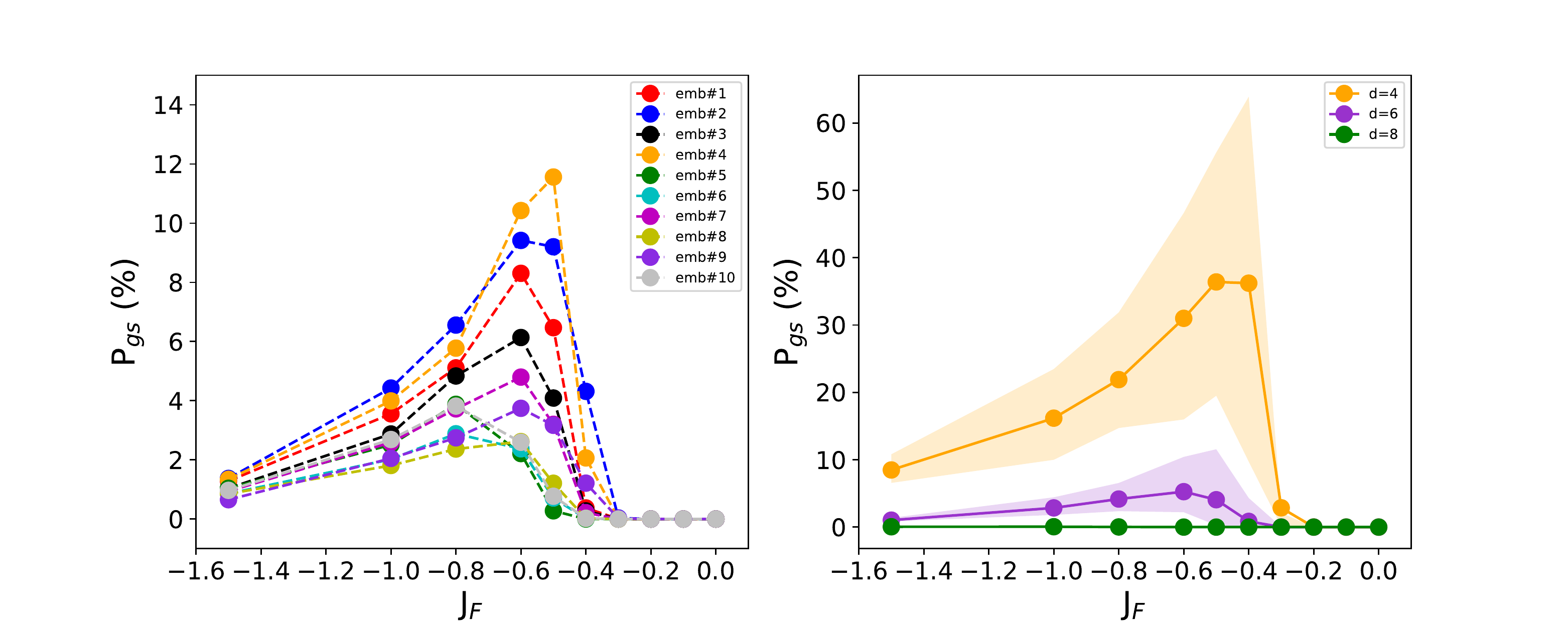}
\caption{Probability of finding the correct ground state (P$_{gs}$) on the D-Wave 2X defined as the ratio between the number of correct solutions and total annealing cycles (40000) as a function of the intra-chain ferromagnetic coupling J$_F$ given in units of the largest coefficient of the embedded Ising model. Only solutions with unbroken chains are counted. Runs are performed using 100 spin reversal operations. Left Panel: We report the case of $d=6$ where different embeddings (different colors) computed with {\tt minorminer} have been considered. Right panel:  Average P$_{gs}$ (solid lines) computed for different problem size $d$. The average is taken over 10 embeddings per problem size. The shaded areas highlight the variation between the lowest and the highest  P$_{gs}$ among the considered embeddings.}
   \label{Jf}
 \end{figure*}

\begin{figure*}[hptb]
  \centering
  \includegraphics[scale=0.15]{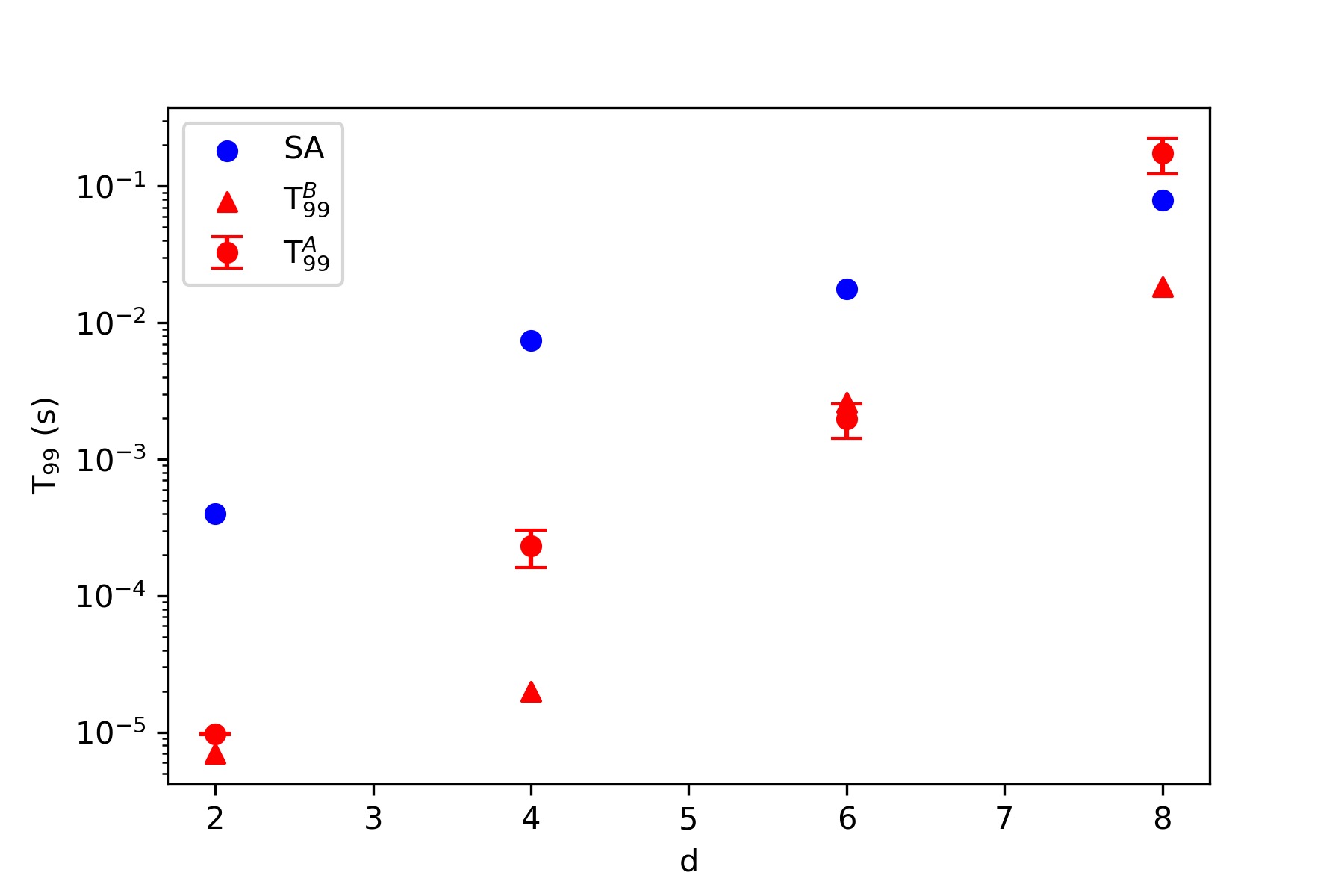}
  \caption{\textcolor{black}{Average time to solution for the default annealing time T=20 $\mu s$ (T$^{A}_{99}$, red dots) and the optimal annealing time for the best performing embedding (T$^{B}_{99}$, red triangles), for different problem sizes ($d$=2,4,6,8).} Data from 10 different embeddings are bootstrapped over 5000 random samples, and the error bars respond $ \pm 2\sigma$ of the bootstrapped sample. In the $d=2$ case, only a single embedding is considered. The scaling of time to solution $T^A_{99}$ with the size $d$ is exponential with coefficients $\alpha_A = 0.52 \pm 0.12$. \textcolor{black}{Blue dots correspond to data obtained with simulated annealing (SA) as implemented in Ocean \cite{Ocean}.} }
  \label{TTS}
\end{figure*}

Fig.\ \ref{Jf} shows the combined effect of $J_F$ and of specific embeddings on P$_{gs}$
for networks with different dimensions ($d=4,6,8$). For $d=6$, the maximum P$_{gs}$ (left panel) is greatly affected both by the choice of the embedding and the optimal $J_F$; that is true for all the considered embeddings. The classical result is matched in 11.6\% of 40,000 annealing cycles. The result for $d=4$ (right panel and Supplementary Material) are similar in term of $J_F$ but the P$_{gs}$ reaches 60\%. For $d=8$, the {optimal $J_F$ is close to 1.0, however, the $J_F$ optimization does not improve significantly the performance in this case.} Indeed, chains are very long ($\geqslant 23$), intra-chain alignment is lost (chain breaking fraction close to one), and the classical solution is practically never reached (0.1\%). We observe that the choice of an optimal $J_F$ improves the performance  and generally depends on the size of the problem. Overall, the choice of the embedding greatly affects the chance of getting a finite value for P$_{gs}$ (see Supplementary Material).\\
The performance in terms of T$^{A}_{99}$ are shown in Fig.\ \ref{TTS} for different sizes of the problem ($d$). As Albash and Lidar have shown before \cite{Tameem_PhysRevX2018}, when D-Wave parameters are not optimized (in their case anneal time), the inferred scaling of the time to solution can be lower than the actual scaling. Not optimizing the embedding will have the same effect because T$_{99}$ can change substantially under different embeddings. How to find an optimal embedding remains an important open question.\\
The solutions accuracy is also influenced by the number of spin reversal (SR) operations which are performed to {limit the effect of persistent bias on the physical couplers}. The observed values for P$_{gs}$ in the absence of spin reversals changes only modestly when we increase SR (see Supporting Material). For the data presented Fig.\ \ref{Jf} and Fig.\ \ref{TTS}, we set the value to SR=100, as in this case it guarantees the convergence of the performance.

\section{Conclusions}
We framed the Chinese postman problem (CPP) as a quadratic unconstrained binary optimization problem and solved it using {\tt qbsolv}, classical simulated annealing, and D-Wave 2X. Further, we used the solution of the CPP to probe local properties of the graph topology.
We devised a simple workflow and explored systematically the parameter setting on the quantum annealer, specifically the effect of embedding, intra-chain coupling $J_F$, spin-reversal, and problem complexity on the quality of the solutions obtained. By analyzing networks with increasing number of odd degree nodes ($d$), we observed the critical role of the embedding and the appropriate scaling of the qubits interactions, $J_F$.
Namely, even in cases where the correct solution is hardly reachable under the default settings, one must tune these knobs to get the solutions and optimize the performance. The efficiency of the quantum annealing has been compared with classical simulated annealing in order to assess any possible quantum advantage, finding that quantum annealing as implemented in D-Wave 2X is on average an order of magnitude faster than classical simulated annealing on commercial traditional hardware.
Although the CPP is solvable classically in polynomial time, generalizations such as the Mixed CPP or the Windy CPP are NP-hard problems which can take advantage from adiabatic quantum optimization. Such generalizations are potential extensions of this work.

\bibliographystyle{spphys}

\end{document}


\title{Supplementary Material: Investigating the Chinese Postman Problem on a Quantum Annealer}

\author{Ilaria Siloi \and Virginia Carnevali \and Bibek Pokharel \and Marco Fornari \and Rosa Di Felice
}
\institute{Ilaria Siloi$^\dagger$  \at
           Department of Physics and Astronomy, University of Southern California, Los Angeles, CA 90089, United States
           \and
           Virginia Carnevali$^\dagger$   \at
           Department of Physics, Central Michigan University, Mt. Pleasant, MI 48859, United States
           \and
           Bibek Pokharel \at
          Department of Physics and Astronomy, University of Southern California, Los Angeles, CA 90089, United States\\
          Center for Quantum Information Science \& Technology, University of Southern California, Los Angeles, California 90089, United States
           \and
          \and
          Marco Fornari$^*$  \at
           Department of Physics and Science of Advanced Materials Program, Central Michigan University, Mt. Pleasant, MI 48859, United States
           \and
          Rosa Di Felice \at
          Department of Physics and Astronomy, University of Southern California, Los Angeles, CA 90089, United States \\
          Center for Quantum Information Science \& Technology, University of Southern California, Los Angeles, California 90089, United States
\and
          $^\dagger$ These Authors contributed equally to this work.\\
          $^*$ Corresponding Author \email{forna1m@cmich.edu}
}


\maketitle



\section{Quadratic Unconstrained Binary Optimization for the Chinese Postman Problem}

The goal of this section is to justify the quadratic unconstrained binary optimization (QUBO) procedure for the Chinese postman problem (CPP) and show the validity of the functional form used in our calculations.
As discussed in Sec. 2 and Sec. 3, the overall quadratic function to be minimized is
\begin{equation}
  Q(\mathbf{x})=\sum_{i\ne j}^{d-1}W_{ij}x_{ij}^2+p P_1(\mathbf{x})+p P_2(\mathbf{x})
  \label{Qubo}
\end{equation}
with
\begin{equation}
  P_1(\mathbf{x})=\sum_{i\ne j}^{d-1}\left(1-\sum_{j=0}^{d-1}(x_{ij}+x_{ji})\right)^2
  \label{P1}
\end{equation}
and
\begin{equation}
  P_2(\mathbf{x})=\sum_{i\ne k; j\ne k}^{d-1}(x_{ik}x_{jk}+x_{ki}x_{kj})
  \label{P2}
\end{equation}
for some constant $p \geqslant d$.
Following Dinneen \textit{et al}. \cite{CS}, we show that (1) the penalty functions, $P_1$ and $P_2$, ensure a legal combination of pairs of vertices with odd degree nodes, and (2) a solution of the variational problem, $\bf{x}_*$, must be a legal combination of pairs of $d$ vertices with odd degree such that $Q(\mathbf{x})=\min_{\alpha}m(\pi_{\alpha})$ with $\alpha$ running over the $(d-1)!!$ perfect matchings.

\begin{itemize}
\item[(1):] A legal combination of pairs of odd degree is a vector $\mathbf{x}\in \mathbb{Z}_{2}^{d(d-1)}$ of decision variables $x_{ij}$ if and only if $P_1(\mathbf{x})= P_2(\mathbf{x})=0$.

\textbf{Proof:} A specific legal $\mathbf{x}$ represents one among the $(d-1)!!$ perfect matchings formed by list of edges between nodes with odd degree. A perfect matching of a graph is a set of pairwise non-adjacent edges such that no two edges share a common vertex and that matches all vertices of the graph: $\pi_{\alpha} \subset V \times V$ such that if $(v_i,v_j)$ and $(v_n,v_m) \in \pi_{\alpha}$ it follows that $i \ne j \ne n \ne m$. We use the cooncept of perfect matching in $G_O(V_O,E_O)$ (the subgraph formed with the vertices with odd degree). This imposes that $x_{ij} = 1 \implies x_{ji} = 0$ and $x_{ik}x_{jk} + x_{ki}x_{kj} = 0\ \forall i\ne k \mbox{ and } \forall j \ne k$. The first condition is equivalent to  $P_1(\mathbf{x})=0$ and the second $P_2(\mathbf{x})=0$.

\item[(2):]$\mathbf{x_*}$ is an optimal variable assignment for $Q(\mathbf{x})$ if and only if $\mathbf{x_*}$ is a legal combination of pairs of nodes with odd degree such that $Q(\mathbf{x_*})= \min_{\alpha}m(\pi_{\alpha})$.

\textbf{Proof:} Given a legal combination of vertices of odd degree (one among the all perfect matchings $\pi_{\alpha}$ in $G_O(V_O,E_O)$  there is a variational mapping, $W$,  which assigns to each element  $(v_i,v_j) \in \pi_{\alpha}$ a weight $W(v_i,v_j)=W_{ij}$ by selecting the minimum among all the possible arcs connecting $v_i$ and $v_j$ in the graph $G(V,E)$ which includes $G_O(V_O,E_O)$:
$$
W(v_i,v_j) = \min\{w(v_i,v_{k_1})+ w(v_{k_1},v_{k_2})+ \dots + w(v_{k_n},v_j))\}
$$
with $v_i,v_j \in V_O$ and $v_{k_l} \in V$.
Suppose that $\mathbf{x}$ is a legal combination of pairs of nodes with odd degree such $Q(\mathbf{x})=\min_{\alpha} m(\pi_{\alpha}) = \min \sum_{(v_i,v_j) \in \pi_{\alpha}} W(v_i,v_j)$. From the previous observation, $P_1(\mathbf{x})=0$ and $P_2(\mathbf{x})=0$. The only relevant term in the $Q(\mathbf{x})$ is
$\sum_{i\ne j}^{d-1}W_{ij}x_{ij}^2 = \sum_{i\ne j}^{d-1}W_{ij}x_{ij} = m(\pi_{\alpha})$.
Finding the minimum $\mathbf{x_*}$ of $Q(\mathbf{x})$ is then equivalent to find the $\min_{\alpha} m(\pi_{\alpha})$.
\end{itemize}

\begin{sidewaystable}[htbp] \centering
\begin{adjustbox}{max width=\textwidth}
  \begin{tabular}{c|ccccccccccc} \toprule
    {$d$} &\begin{tabular}[c]{@{}c@{}}P$_{gs}$ \\ all chains \\(\%)\end{tabular} & \begin{tabular}[c]{@{}c@{}}P$_{gs}$ \\ unbroken only \\(\%)\end{tabular}& \begin{tabular}[c]{@{}c@{}}number of \\ physical\\ qubits\end{tabular}& \begin{tabular}[c]{@{}c@{}}T$_{99}$ \\ all chains \\(s)\end{tabular} & \begin{tabular}[c]{@{}c@{}}T$_{99}$ \\ unbroken only \\(s)\end{tabular} & \begin{tabular}[c]{@{}c@{}}max chain\\ length\end{tabular} & \begin{tabular}[c]{@{}c@{}}number of\\ chain with \\ max length\end{tabular} & \multicolumn{4}{c} {eccentricity } \\\cline{9-12}
      &       &        &   &         &          &   &     & mean & variance & skewness & kurtosis \\ \midrule
    4 & 52.16 & 22.77 & 44 & 1.25e-4 & 3.56e-4  & 5 & 1  & 5.25  & 0.64 & 0.05   & -0.63    \\
    4 & 38.86 & 38.68 & 44 & 1.87e-4  & 1.88e-4   & 4 & 8  & 5.52  & 0.48 & 0.13   & -0.25   \\
    4 & 19.82 & 19.68 & 45 & 4.17e-4  & 4.20e-4   & 5 & 2  & 5.29   & 0.56 & 0.11   & -0.34  \\
    4 & 68.45 & 55.62 & 44 & 7.98e-5   & 1.13e-4  & 5 & 2  & 5.32   & 0.58 & 0.01  & -0.45   \\
    4 & 31.77 & 31.61 & 47 & 2.41e-4 & 2.42e-4  & 5 & 3  & 5.49    & 0.55 & 0.19   & -0.30  \\
    4 & 51.37 & 50.86 & 44 & 1.28e-4  & 1.30e-4  & 5 & 2  & 5.41   & 0.51 & -0.05 & -0.31  \\
    4 & 25.16 & 21.11 & 45 & 3.18e-4  & 3.88e-4   & 5 & 2  & 5.44   & 0.47 & 0.41   & -0.10  \\
    4 & 87.83 & 45.05 & 44 & 4.37e-5  & 1.54e-4  & 5 & 1  & 5.57   & 0.43 & 0.72    & -0.53   \\
    4 & 70.85 & 63.91 & 45 & 7.47e-5   & 9.04e-5   & 5 & 1  & 5.07   & 0.55 & -0.11  & -1.18   \\
    4 & 31.68 & 31.42 & 46 & 2.42e-4  & 2.44e-4   & 5 & 2  & 5.46   & 0.51 & 0.15    & -0.23  \\ \midrule
    6 & 51.00 &  8.30 & 248 & 1.29e-4 & 1.06e-3  & 11 & 2 & 10.34 & 1.51 & 0.07 & -0.60    \\
    6 & 16.68 &  9.41 & 262 & 5.04e-4 & 9.31e-4 & 12 & 1 & 11.31 & 2.07 & 0.03 & -0.72   \\
    6 &  7.48 &  0.61 & 274 & 1.18e-4 & 1.45e-3 & 13 & 1 & 11.45 & 2.22 & 0.04 & -0.56  \\
    6 & 15.18 & 11.56 & 257 & 5.59e-4 & 7.50e-4 & 11 & 1 &10.38 & 1.79 & -0.07 & -0.67   \\
    6 &  8.12 &  3.88 & 284 & 1.09e-3 & 2.32e-3 & 13 & 1 & 11.63 & 2.33 & -0.06 & -0.54  \\
    6 & 14.76 &  2.88 & 291 & 5.76e-4 & 3.15e-3 & 13 & 3 & 11.92 & 2.26 & 0.10 & -0.65  \\
    6 &  5.82 &  4.80 & 261 & 1.54e-3 & 1.87e-3 & 12 & 1 & 10.56 & 1.80 & -0.05 & -0.71  \\
    6 &  3.37 &  2.62 & 274 & 2.68e-3 & 3.47e-3 & 12 & 2 & 11.20 & 1.85 & -0.02 & -0.68   \\
    6 &  4.55 &  3.74 & 272 & 1.98e-3 & 2.41e-3 & 13 & 1 & 10.99 & 2.05 & -0.13 & -0.64  \\
    6 & 10.53 &  3.81 & 257 & 8.27e-4 & 2.37e-3 & 11 & 3 & 11.06 & 2.35 & 0.12 & -0.44  \\ \midrule
    8 & 0.21 & 0.09 & 864 & 4.38e-2 & 9.69e-2 & 23 & 1 & 18.94 & 6.07 & -0.01 & -0.57    \\
    8 & 0.20 & 0.06 & 901 & 4.66e-2 & 1.47e-1 & 24 & 1 & 19.43 & 5.96 & -0.01 & -0.60  \\
    8 & 0.19 & 0.06 & 879 & 4.78e-1 & 1.53e-1 & 22 & 3 & 19.01 & 5.73 & 3e-4 & -0.69\\
    8 & 0.12 & 0.04 & 902 & 7.36e-2 & 2.17e-1 & 23 & 2 & 18.67 & 5.87 & -0.07 & -0.62 \\
    8 & 0.16 & 0.06 & 896 & 5.66e-2 & 1.36e-1 & 26 & 1 & 18.96 & 5.73 & -0.08 & -0.55  \\
    8 & 0.18 & 0.10 & 854 & 5.11e-2 & 9.21e-2 & 23 & 1 & 18.92 & 5.54 & 0.06 & -0.69 \\
    8 & 0.12 & 0.04 & 899 & 7.36e-2 & 2.30e-1 & 24 & 1 & 19.08 & 5.86 & 4e-3 & -0.58  \\
    8 & 0.17 & 0.02 & 869 & 5.33e-1 & 3.68e-1 & 23 & 1 & 18.82 & 5.77 & -0.05 & -0.59   \\
    8 & 0.15 & 0.05 & 934 & 6.13e-2 & 1.84e-1 & 25 & 1 & 19.39 & 5.91 & -2e-3 & -0.55  \\
    8 & 0.19 & 0.08 & 884 & 4.72e-2 & 1.19e-1 & 25 & 1 & 18.92 & 5.97 & -0.11 & -0.60 \\
    \bottomrule
  \end{tabular}
  \end{adjustbox}
\caption{Parameters and performance of all the embeddings for the CPP cases studied with
D-Wave 2X: number of nodes with odd degree (d), best probability of finding the correct ground state P$_{gs}$ considering all the chain, best probability of finding the correct ground state P$_{gs}$ considering only the unbroken chains, number of physical qubits required in the embedding, best time to solution T$_{99}$ considering all the chains, best time to solution T$_{99}$ considering only the unbroken chains, maximum chain length of the embedding, number of the chains that have maximum length in the embedding; eccentricity mean, variance, skewness, and kurtosis of the embedding. The embeddings are generated with {\tt minorminer}.}
  \label{TabDW2}
\end{sidewaystable}

\newpage
\section{Results by Quantum Annealing: with and without majority voting}
\begin{figure*}[hptb]
  \centering
\includegraphics[scale=0.35]{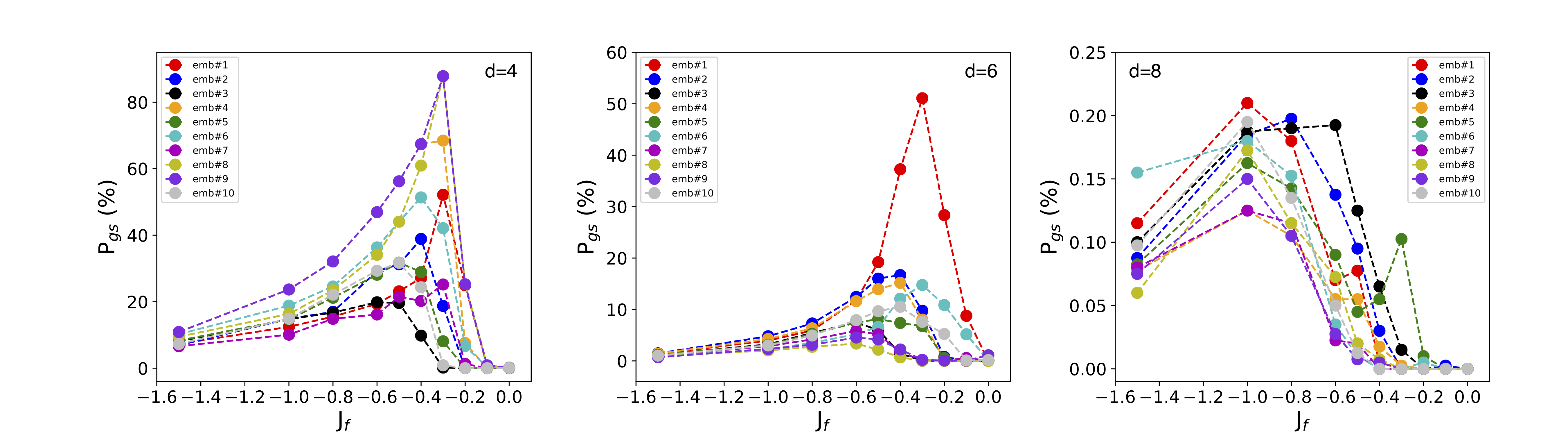}
  \caption{Probability of finding the correct ground state (P$_{gs}$) on the D-Wave 2X as a function of the intra-chain ferromagnetic coupling $J_F$ given in units of the largest coefficient of the embedded Ising model. We report all the correct solutions including broken chains that are post-processed using majority-voting. Runs are performed using 100 spin reversal operations. Different panel corresponds to different sizes of the problems $d=4$ (left panel), 6 (central panel), 8 (right panel). For each case 10 different embeddings (different colors) have been considered.}
  \label{S1_all}
\end{figure*}

\begin{figure*}[hptb]
  \centering
\includegraphics[scale=0.35]{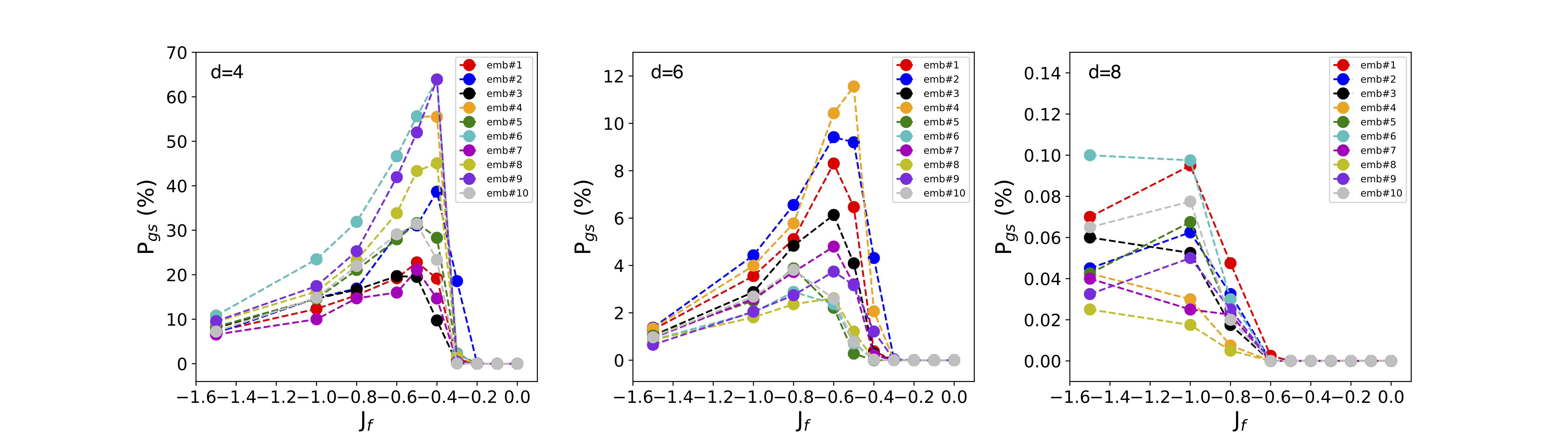}
  \caption{Probability of finding the correct ground state (P$_{gs}$) as a function of the intra-chain ferromagnetic coupling $J_F$ given in units of the largest coefficient of the embedded Ising model. We report all the correct solutions obtained for with unbroken chains. Runs are performed using 100 spin reversal operations. Different panel corresponds to different sizes of the problems $d$=4 (left panel), 6 (central panel), 8 (right panel). For each case 10 different embeddings (different colors) have been considered.}
  \label{S2_ub}
\end{figure*}

In this section we report a comparison in the performance of DW2X, measured as probability of finding the correct ground state (P$_{gs}$), in the case of majority voting as opposed to the use of unbroken chain only (see also Tab.\ \ref{TabDW2}). We observe that the use of a decoding procedure (majority voting) increases substantially the P$_{gs}$ regardless the size of the QUBO problem considered, see Fig.\,\ref{S1_all} and Fig.\,\ref{S2_ub}. Such increment is not related to the ``quality" of the embedding as the ranking of the best embedding is different in the case of majority voting. Interestingly, the use of broken chain allows finite P$_{gs}$ for values of intra-chain coupling that do not count any valid solutions in the case of unbroken chains.  The optimal $J_F$ value does not vary considerably across all the embeddings without decoding, while majority voting complicates this picture, especially $d$=8.

\section{Results by Quantum Annealing: spin reversal transformation}

\begin{figure*}[hptb]
\centering
\includegraphics[scale=0.35]{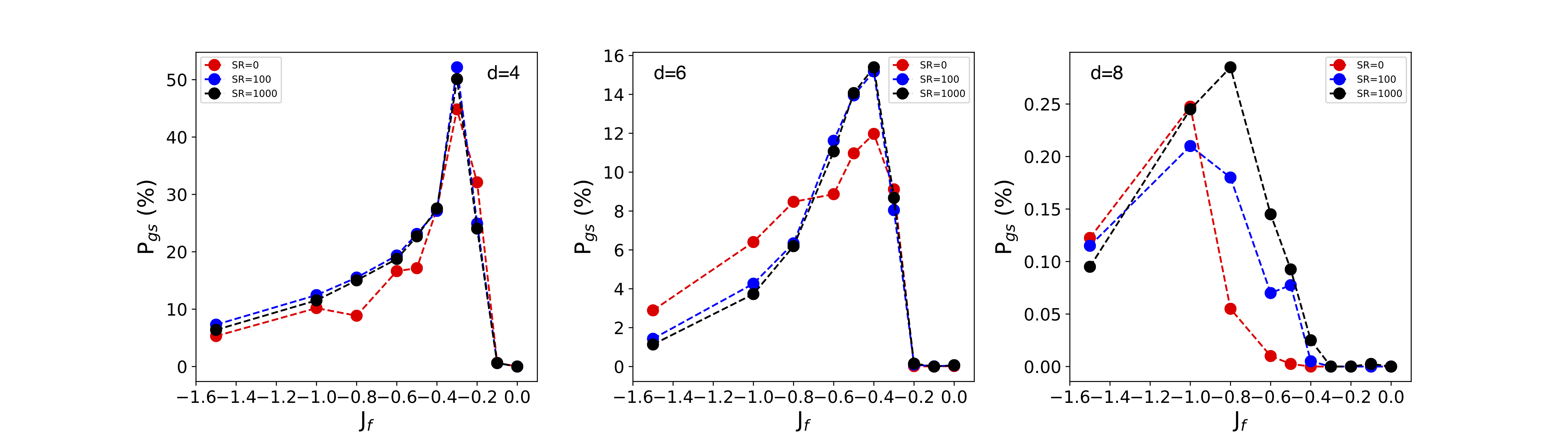}
\caption{P$_{gs}$ as a function of the intra-chain ferromagnetic coupling J$_f$ given in units of the largest coefficient of the embedded Ising model. For each size of the problem,  $d$=4 (left panel), 6 (central panel), 8 (right panel),  only one embedding is shown. Qualitatively similar results are obtained for all the other embeddings. We report all the correct solutions including broken chains that are post-processed using majority-voting. Colors mark a different number of spin reversal operations 0 (red), 100(blue), 1000 (black).}
 \label{S3_all}
 \end{figure*}

\begin{figure*}[hptb]
\centering
\includegraphics[scale=0.35]{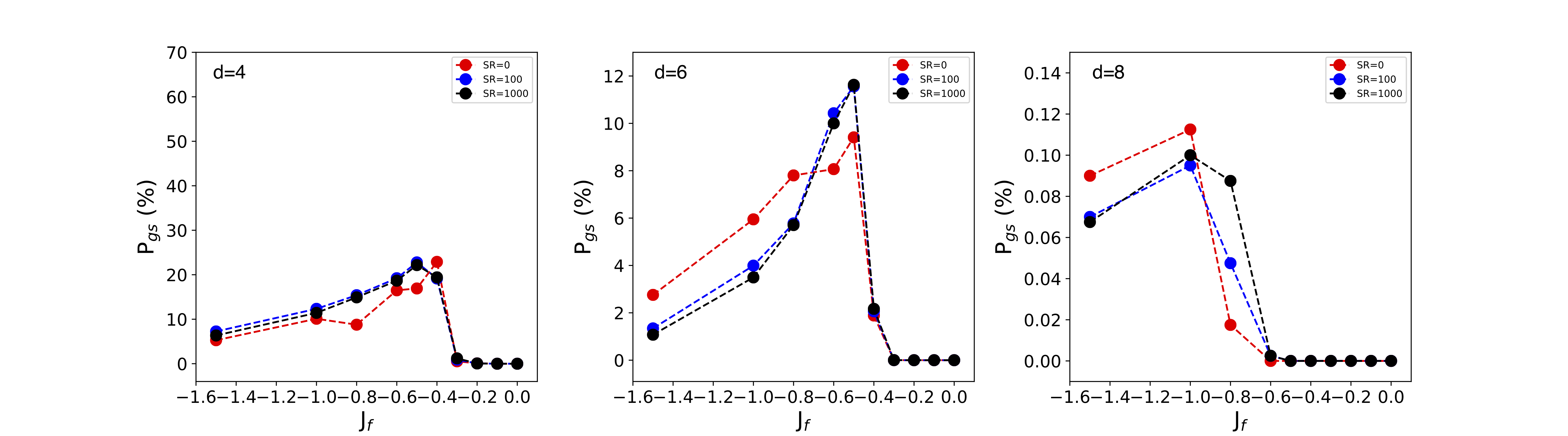}
\caption{P$_{gs}$ as a function of the intra-chain ferromagnetic coupling J$_f$ given in units of the largest coefficient of the embedded Ising model. For each size of the problem,  $d$=4 (left panel), 6 (central panel), 8 (right panel),  only one embedding is shown. Qualitatively similar results are obtained for all the other embeddings. Only unbroken chains are considered. Colors mark a different number of spin reversal operations 0 (red), 100(blue), 1000 (black).}
 \label{S4_ub}
 \end{figure*}

Here we look at the effect of increasing the number of gauges (spin reversal transformations). We focus our analysis on a single embedding as the result does not change qualitatively for the others. Spin reversal transformations are known to increase the accuracy of the calculation. Fig.\,\ref{S3_all} and Fig.\,\ref{S4_ub} show a modest improvement in the convergence of the P$_{gs}$ regardless the decoding procedure. There is no advantage in increasing the number of spin reversal beyond 100.

\section{Comparison between Quantum Annealing, Simulated Annealing, and \tt{Qbsolv}}
\begin{figure*}[hptb]
\centering
\includegraphics[scale=0.9]{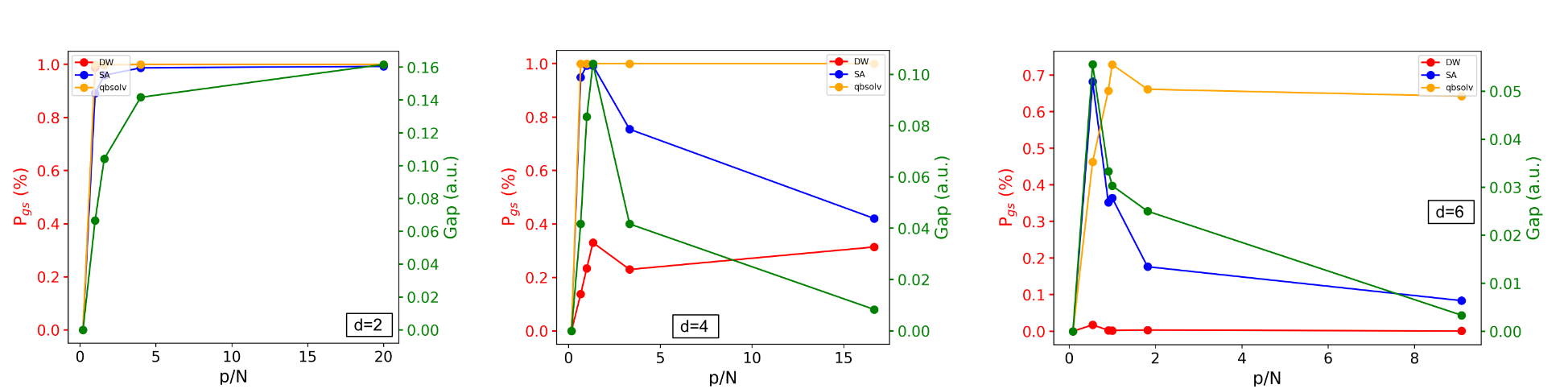}
\caption{P$_{gs}$ as a function of the penalty function $p$ divided by the number of nodes of the graph $N$. For each size of the problem,  $d$=2 (left panel), 4 (central panel), 6 (right panel), the P$_{gs}$ obtained with DW2X (red), simulated annealing (blue), and qbsolv (yellow) have been plotted. In green, gap between the ground state and the first excited state of the QUBO.}
 \label{pgs_p}
 \end{figure*}

In this section, we report a comparison in the performance, measured as probability of finding the correct ground state (P$_{gs}$),
between DW2X, simulated annealing (SA), and {\tt qbsolv}. The P$_{gs}$ has been computed for different value of the penalty $p$ (Eq.\ref{Qubo}) and for each method. The coefficients in front of the penalty functions are known to have a main role in tuning the gap between the ground state and the first excited state of the QUBO. Indeed, we notice that the P$_{gs}$ for quantum annealing and simulated annealing decreases with the increasing of the gap (Fig.\ref{pgs_p}). This is also true for {\tt qbsolv} for $d=6$, while in the case of lower size of the problem, it always finds the solution.

\bibliographystyle{spphys}
\bibliography{mybibfile}